\documentclass[a4paper,12pt]{article}
\usepackage{jheppub}
\usepackage{graphicx}
\usepackage{mathtools}
\usepackage{array}
\usepackage{amsfonts}
\usepackage[utf8]{inputenc} 
\usepackage{amsmath}
\usepackage{pdfpages}
\usepackage{hepunits}
\usepackage{physics}
\usepackage{bm}
\usepackage{tikz}
\usepackage{dsfont}
\usepackage[english]{babel}

\usepackage{verbatim}
\usepackage[thinlines]{easytable}
\usepackage{dsfont}

\title{Chaos and operator growth in 2d CFT}
\author{Surbhi Khetrapal}
\affiliation{School of Physics, University of Hyderabad,
Gachibowli 500046, Hyderabad, India}
\emailAdd{surbhikhetrapal@uohyd.ac.in}
\abstract{We study the out-of-time-ordered correlator (OTOC) in a zero temperature $2d$ large-$c$ CFT under evolution by a Liouvillian composed of the Virasoro generators. A bound was conjectured in \cite{Parker:2018yvk} on the growth of the OTOC set by the Krylov complexity which is a  measure of operator growth. The latter grows as an exponential of time with exponent $2\alpha$, which sets an upper bound on the Lyapunov exponent, $\Lambda_L \leq 2\alpha$. We find that for a two dimensional zero temperature CFT, the OTOC decays exponentially with a Lyapunov exponent which saturates this bound. We show that these Virasoro generators form the modular Hamiltonian of the CFT with half space traced out. Therefore, evolution by this modular Hamiltonian gives rise to thermal dynamics in a zero temperature CFT. Leveraging the thermal dynamics of the system, we derive this bound in a zero temperature CFT using the analyticity and boundedness properties of the OTOC.}

\begin{document}
\maketitle

\section{Introduction}
Growth of quantum operators in local systems encodes various aspects of quantum dynamics such as the system's transport properties and emergence of quantum chaos.  This growth studied in Krylov basis was proposed as a potential link between out-of-time-ordered correlators, widely used to study quantum chaos in black-holes, and conventional signatures of many-body quantum chaos.

Operator growth in quantum many body systems are studied using a quantity called the Krylov complexity or $K$-complexity, which for a system undergoing Liouvillian (or Hamiltonian) dynamics is the expectation value 
\begin{align}
   (K)_t = \langle {\cal O}(t)|K|{\cal O}(t) \rangle 
\end{align}
of the operator 
\begin{align}
    K = \sum_n n |{\cal O}_n \rangle \langle {\cal O}_n |
\end{align}
where $|{\cal O}_n \rangle$ is the Krylov basis and the operators ${\cal O}(t)$ are evolved under the Liouvillian, ${\cal L}=[H,*]$. The operator growth may be studied under evolution by a Liouvillian which is a Hermitian operator but need not be the usual time evolution Hamiltonian of the theory.

In this paper, we study operator growth in a $2d$ CFT, under a special class of Hermitian operators composed of the Virasoro generators,
\begin{align}\label{Hk}
    {H}_k = \alpha \left( L_k + L_{-k} \right), 
\end{align}
where $k$ belongs to positive integers $k \in \mathbb{Z}^+$.\footnote{It was discussed in \cite{Caputa:2021sib} that the Liouvillian of the form ${\cal L} = H_1$ acts as a Hamiltonian evolution in the Krylov basis in a $2d$ CFT.} Therefore, the operators will evolve by the following unitary,
\begin{align} \label{unitary1}
    { U}_k(s)=e^{i \alpha (L_k+L_{-k})s+\dots}
\end{align}
instead of the usual CFT time evolution Hamiltonian. In the above equation, the dots $\dots$ denote the anti-holomorphic terms which we will not keep track of in this paper. This type of evolution is encountered in a CFT undergoing inhomogenous quench \cite{Das:2022jrr}. Since the evolution is not by the usual CFT Hamiltonian $\propto L_0$, here $s$ can be thought of as a notion of internal time of the operator as it grows under the action of this unitary,
\begin{align}
    { U}_k(s) V(0) { U}_{k}^\dagger(s) = V(s).
\end{align}
The Krylov complexity of an operator undergoing evolution by the unitary ${ U}_1$ is known \cite{Caputa:2021ori},
\begin{align}
    K_V \approx e^{2\alpha  t}, \quad t \to \infty.
\end{align}
Similarly, for an operator undergoing evolution by ${ U}_k$ where $k=\{2,3,\dots\}$, the Krylov complexity \cite{Caputa:2021sib} is $K_V \approx e^{2\alpha k t}$ for $t \to \infty$.

$K$-complexity and operator growth have been used as a probe of information scrambling in many-body quantum systems. It was shown that the growth rate of $K$-complexity imposes an upper-bound on the Lyapunov exponent of the theory. This was proven for an infinite temperature theory and conjectured for a finite temperature theory in \cite{Parker:2018yvk}. Following this work, there has been interest in understanding this bound \cite{Avdoshkin:2019trj,Dymarsky:2021bjq,Gu:2021xaj} under evolution by the CFT Hamiltonian. It was conjectured in \cite{Parker:2018yvk}, that in a quantum system, the rate of growth of Krylov complexity, i.e. $2\alpha$, sets an upper bound on the rate of growth of the out-of-time ordered correlator (OTOC). Thus the Lyapunov exponent of the OTOC is expected to be bounded as,
\begin{align}
    \Lambda_L \leq 2\alpha.
\end{align}

In this work, we find that in a 2d CFT at zero temperature, undergoing time evolution by the Hermitian operators of equation \eqref{Hk}, the Lyapunov exponent characterising the OTOC growth saturates this bound. Since we find exponential growth of the OTOC in a zero temperature CFT, this indicates that evolution by the operator in equation \eqref{Hk} gives rise to thermal dynamics. This thermal dynamics comes about since the Hermitian operators ${H}_k$ are modular Hamiltonians with half space traced out. Half-space modular Hamiltonians annihilate the CFT vacuum, give rise to local dynamics and satisfy the KMS conditions \cite{cmp/1104253848, Balakrishnan:2017bjg}. 

Leveraging the thermal nature of the system, we derive the bound on Lyapunov exponent set by Krylov complexity from an argument similar to \cite{Maldacena:2015waa} and which uses the analyticity and boundedness properties of the OTOC. The fact that this bound is saturated hints at the existence of a holographic theory with a black hole of inverse temperature $\beta = \frac{\pi}{\alpha}$. 

In section \ref{sec:OTOC}, we compute the OTOC under evolution by the unitary ${U}_1$ first using vacuum dominance and then including light intermediate exchanges. In both the cases, we find that the OTOC exponentially decays with the Lyapunov exponent $\Lambda_L = 2\alpha$. We also derive the scrambling time as a function of $\alpha$. We go on to compute the OTOC when the evolution is by the unitary ${U}_k$ for integer $k>0$. Using this, we conclude that the upper bound set on the Lyapunov exponent by the rate of growth of $K$-complexity is saturated under operator evolution by all Virasoro algebra generators. In section \ref{sec:modular_hamiltonian}, we start with a brief review of modular Hamiltonians and go on to show that the Hamiltonian ${H}_k$ is the modular Hamiltonian with half-space of the CFT traced out. In section \ref{sec:bound}, we show that the bound on the Lyapunov exponent obtained by exploiting the analyticity and boundedness properties of the OTOC is the same as the bound proposed in \cite{Parker:2018yvk}. In this paper we have shown this bound for a zero temperature CFT but our arguments may be generalised for a finite temperature CFT.

%\section{OTOC in 2d CFT under Liouvillian evolution} 
\section{OTOC under evolution by Virasoro generators}\label{sec:OTOC}
To compute the OTOC, we need to evaluate a $4$-point correlator of a pair of operators evolved by the unitary $U_1(t) = e^{i\alpha(L_1+L_{-1})t}$,
\begin{align}
     U_1(t) V(z_3)   U_1^\dagger(t)    U_1(t) V(z_4)  U_1^\dagger(t) |0 \rangle  =  U_1(t) V(z_3) V(z_4) |0 \rangle,
\end{align}
with respect to a pair of probe operators. Here, $ U_1^\dagger(t)    U_1(t) = \mathbb{I}$ and $U_1^\dagger(t) |0 \rangle = |0 \rangle$ has been used, and $t$ is a parameter of evolution under the unitary, ${U}_1$. $t$ is a notion of time in the internal clock of the operator, this will be explained further in section \ref{sec:modular_hamiltonian}. The unitary acts on the operators as an $SL(2,R)$ transformation and results in M\"{o}bius transformation of the location of the operators as (see derivation in appendix \ref{app:sl2r}),
\begin{align} \label{z_transform_U}
    z' = \frac{z \cosh(\alpha t) + i \sinh (\alpha t) }{\cosh(\alpha t) - i z \sinh (\alpha t) }.
\end{align}
The growth of a product of operators under the action of this unitary is,
\begin{align}
       U_1(t) V(z_3)   U_1^\dagger(t)    U_1(t) V(z_4)  U_1^\dagger(t) |0 \rangle = \left( \frac{\partial z'}{\partial z} \right)^{h_v}\bigg|_{z=z_3} \left( \frac{\partial z'}{\partial z} \right)^{h_v}\bigg|_{z=z_4} V(z'_3) V(z'_4) |0 \rangle,
\end{align}
where $z'$ are given by equation \eqref{z_transform_U}.

Using the above transformation, the $4$-point correlator where two operators are evolved in time and the two probe operators are held fixed is,
\begin{align} \label{4pt_u1}
    \langle 0| & W(z_1) W(z_2)     U_1(t) V(z_3)   U_1^\dagger(t)    U_1(t) V(z_4)  U_1^\dagger(t) |0 \rangle \nonumber \\
    &= \left( \frac{\partial z'}{\partial z} \right)^{h_v}\bigg|_{z=z_3} \left( \frac{\partial z'}{\partial z} \right)^{h_v}\bigg|_{z=z_4} \langle 0| W(z_1) W(z_2) V(z'_3) V(z'_4) |0 \rangle. 
\end{align}
So as not to keep track of the factors in front, we will work with the normalised correlator in the rest of the paper,
\begin{align}
    {\cal C} & = \frac{ \langle 0| W(z_1) W(z_2)    U_1(t) V(z_3)   U_1^\dagger(t)    U_1(t) V(z_4)  U_1^\dagger(t) |0 \rangle}{\langle 0| W(z_1) W(z_2) |0\rangle \langle 0|    U_1(t) V(z_3)   U_1^\dagger(t)    U_1(t) V(z_4)  U_1^\dagger(t) |0 \rangle} \nonumber \\
    & = \frac{\langle 0| W(z_1) W(z_2) V(z'_3) V(z'_4) |0 \rangle}{\langle 0| W(z_1) W(z_2) |0 \rangle \langle 0| V(z'_3) V(z'_4) |0 \rangle}.
\end{align}

To obtain the OTOC from the above correlator we use the prescription in \cite{Craps:2021bmz,Roberts:2014ifa}. First, the operators are placed at some location $z=\sigma + i \tau$ in the complex $z$ plane, $\tau$ is the Euclidean time. Next, assign a small imaginary time to each operator location $\tau = \epsilon$ such that $z_j = \sigma_j+i \epsilon_j$. Thus the operator locations in the $z$ plane are:
\begin{align} \label{z_locations}
    z_1  =i \epsilon_1,\quad  z_2=i \epsilon_2, \quad  z_3=\sigma + i \epsilon_3, \quad z_4=\sigma + i \epsilon_4,\\
     \bar{z}_1 = -i \epsilon_1,\quad  \bar{z}_2 = -i \epsilon_2, \quad  \bar{z}_3 = \sigma - i \epsilon_3, \quad \bar{z}_4 = \sigma - i \epsilon_4.
\end{align}
The $4$-point correlator is out-of-time ordered for the following operator ordering in imaginary time,
\begin{align} \label{operator_ordering}
    \epsilon_1<\epsilon_3<\epsilon_2<\epsilon_4.
\end{align}
Finally, to obtain the OTOC, keeping the imaginary time fixed, the operators are evolved with the unitary, ${U}_1$. The locations of the evolved operators $z'_{3,4}$ depend on time $t$ and are obtained by substituting $z_{3,4}$ from \eqref{z_locations} in \eqref{z_transform_U}. Thus, due to the out-of-time ordering of the operators in imaginary time given by equation \eqref{operator_ordering}, the $4$-point correlator becomes following OTOC with appropriate normalisation,
\begin{align}
    \langle 0|W(z_1)  U_1(t) V(z_3) U_1^\dagger (t)   W(z_2) U_1(t) V(z_4) U_1^\dagger(t) |0\rangle.
\end{align}

To evaluate the above OTOC, the Virasoro block decomposition of 4-point correlation functions is used. Thus the normalised correlator is
\begin{align} \label{block_expansion}
{\cal C} = \sum_{h,\bar{h}} P_{h,\bar{h}}{\cal V}_{h,\bar{h}}(u,v),
\end{align}
where, ${\cal V}_{h,\bar{h}}$ are the Virasoro blocks and $P_{h,\bar{h}}$ are the OPE coefficients. The variables $u$ and $v$ are functions of the conformal cross-ratios, $\eta$ and $\bar{\eta}$,
\begin{align}
u = \eta \, \bar{\eta} , \qquad v=\left( 1-\eta \right)\left( 1- \bar{\eta} \right),
\end{align}
where, the conformal cross-ratios are
\begin{align} \label{eq:cross-ratio_def}
\eta = \frac{(z_1-z_2)(z'_3-z'_4)}{(z_1-z'_3)(z_2-z'_4)}, \qquad \bar{\eta} = \frac{(\bar{z}_1-\bar{z}_2)(\bar{z}'_3-\bar{z}'_4)}{(\bar{z}_1-\bar{z}'_3)(\bar{z}_2-\bar{z}'_4)}.
\end{align}
Substituting equations \eqref{z_locations} and \eqref{z_transform_U} in above, we get the cross-ratio,
\begin{align}
    \eta = \frac{(\epsilon_1-\epsilon_2) (\epsilon_3-\epsilon_4)}{(\sinh (\alpha  t)-i \sigma  \cosh (\alpha  t))^2}+O\left(\epsilon ^3\right) 
\end{align}
and similarly for $\bar{\eta}$.

\subsection{Vacuum block dominance}
The next step is to use the Virasoro identity block approximation, which means that the correlator is approximated by the contribution of the Virasoro block of the identity operator, ${\cal V}_{0,0}(u,v)$. This provides a good approximation to semi-classical gravity in AdS$_3$. Furthermore, it was shown in \cite{Fitzpatrick:2014vua} that the Virasoro blocks factorise,
\begin{align} \label{virasoro_block_factorise}
{\cal V}_{0,0}(u,v) = {\cal V}_0(\eta) {\cal V}_0(\bar{\eta}),
\end{align}  
where ${\cal V}_0(\eta)$ and ${\cal V}_0(\bar{\eta})$ are purely holomorphic and anti-holomorphic components, respectively. 
We consider the vacuum block in the semi-classical limit, $c \to \infty$, where $\frac{h_v}{c} \ll 1$, $\frac{h_p}{c} \ll 1$ while keeping $\frac{h_w}{c}$ arbitrary (also called the heavy-heavy-light-light (HHLL) limit), 
\begin{align}
    {\cal V}_0(\eta) = \left( \frac{\eta}{1-(1-\eta)^{1-\frac{12h_w}{c}}}\right)^{2h_v}.
\end{align}
It can be seen that this function has a branch cut from $\eta = [0,\infty)$, where $\eta$ is a complex variable. 

With $\sigma$ held fixed as the time is increased, for the operator ordering \ref{operator_ordering} the branch cut is crossed in the complex $\eta$ plane 
\begin{align}\label{eta_continue}
    (1-\eta) \to (1-\eta) e^{-2 \pi i}.
\end{align}
Thus the block becomes in the $\eta \sim 0$ limit,
\begin{align}
    {\cal V}_0(\eta) = \left( 1-\frac{24\pi i h_w}{c\,\eta}\right)^{-2h_v},
\end{align}
which is the relevant limit for computing the OTOC at late times. The cross-ratio approaches zero, $\eta \to 0$, in the large $t$ limit,
\begin{align}\label{cross-ratio}
    \eta = -\frac{4 e^{-2 \alpha  t} (\epsilon_1-\epsilon_2) (\epsilon_3-\epsilon_4)}{(\sigma+i )^2}.
\end{align}
Substituting this limit of the cross-ratio in the block, we obtain the following OTOC:
\begin{align} \label{otoc_u1}
    {\cal C} & = \left( 1+\frac{6\pi i h_w }{c \, \epsilon_{12}\epsilon_{34}}(\sigma+i)^2  e^{2 \alpha  t}\right)^{-2h_v}\\
    & \approx 1 - \frac{12\pi i h_w h_v }{c \, \epsilon_{12}\epsilon_{34}}(\sigma+i)^2  e^{2 \alpha  t}
\end{align}
Thus the Lyapunov exponent is read-off to be the co-efficient of $t$ in the exponential, and here it is $\Lambda_L=2\alpha$, which saturates the bound proposed in \cite{Parker:2018yvk} that $\Lambda_L \leq 2\alpha$.

\subsection{Light intermediate states}
In the previous sub-section, we computed the OTOC by restricting to the identity block approximation. However, in this section we compute the contribution to the OTOC of the subleading terms in the conformal block expansion of equation \eqref{block_expansion}. This will provide a check of whether the bound $\Lambda_L \leq 2\alpha$ is saturated even without restricting to the identity block approximation. Let's focus on the contribution of the following term,
\begin{align} \label{light_block}
    {\cal V}_{h_p,\bar{h}_p}(u,v) =  {\cal V}_{h_p}(\eta)  {\cal V}_{\bar{h}_p}(\bar{\eta}),
\end{align}
in the semi-classical limit, $c \to \infty$.
In this limit, taking $\frac{h_v}{c} \ll 1$, $\frac{h_p}{c} \ll 1$ while keeping $\frac{h_w}{c}$ arbitrary, the function ${\cal V}_{h_p}(\eta)$ is \cite{Fitzpatrick:2014vua},
\begin{align}
    {\cal V}_{h_p}(\eta) = \left[\frac{\gamma \eta (1-\eta)^{(\gamma-1)/2}}{1-(1-\eta)^\gamma}\right]^{2h_v} \left[ \frac{4}{\gamma}\frac{(1-(1-\eta)^{\gamma/2})}{(1+(1-\eta)^{\gamma/2})}\right]^{h_p}.
\end{align}
where,
\begin{align}
    \gamma = \sqrt{1-\frac{24h_w}{c}}.
\end{align}

The contribution to the OTOC is obtained by analytically continuing across the brach cut \eqref{eta_continue} followed by taking $\eta \ll 1$ and $h_w/c \ll 1$ limits,
\begin{align}
     {\cal V}_{h_p}(\eta) & = \left[\frac{ \eta }{1- e^{-2\pi i \gamma}(1-\eta)}\right]^{2h_v} \left[ \frac{4 (1- e^{-\pi i \gamma} (1-\eta)^{1/2})}{(1+ e^{-\pi i \gamma} (1-\eta)^{1/2})}\right]^{h_p}, \nonumber\\
     & = \left[\frac{ \eta }{1- e^{\frac{24 \pi i h_w}{c}}(1-\eta)}\right]^{2h_v} \left[ \frac{4 (1+ e^{\frac{12 \pi i h_w}{c}} (1-\eta)^{1/2})}{(1- e^{\frac{12 \pi i h_w}{c}} (1-\eta)^{1/2})}\right]^{h_p} \nonumber\\
     & = \left( \frac{1}{1-\frac{24 \pi i h_w}{c \eta }} \right)^{2h_v+h_p} \left( \frac{16}{\eta}\right)^{h_p}
\end{align}
where we have substituted $\gamma \approx 1-\frac{12h_w}{c}$ in the second step and further simplified the equation in the last step. The above block has different behaviour in the following three regimes \cite{Hampapura:2018otw},
\begin{align}
    {\cal V}_{h_p}(\eta) = \begin{cases}
    \left( \frac{16}{\eta}\right)^{h_p}, & \frac{h_w}{c} \ll \eta\\
    \left( \frac{1}{1-\frac{24 \pi i h_w}{c \eta }} \right)^{2h_v+h_p} \left( \frac{16}{\eta}\right)^{h_p}, & \frac{h_w}{c} \sim \eta\\
    16^{h_p} \left(\frac{i \, c}{24\pi h_w}\right)^{2h_v+h_p} \eta^{2h_v}, & \frac{h_w}{c} \gg \eta.
    \end{cases}
\end{align}

We are restricting to the channel \cite{Craps:2021bmz} where the anti-holomorphic cross-ratio does not cross the branch-cut, therefore for all time it remains,
\begin{align}
    {\cal V}_{\bar{h}_p}(\bar{\eta}) = \bar{\eta}^{\bar{h}_p}.
\end{align}
Substituting the cross-ratio $\eta$ from equation \eqref{cross-ratio}, we get the following contribution of the light intermediate states to the OTOC,
\begin{align}
    & {\cal V}_{h_p}(\eta) {\cal V}_{\bar{h}_p}(\bar{\eta}) = \nonumber\\
   & \begin{cases}
    16^{h_p} \left(-\frac{ \epsilon_{12} \epsilon_{34}}{\sigma ^2}\right)^{\bar{h}_p-h_p}  \left(1+\frac{\alpha ^2t^2 }{\sigma ^2} \left(  \left(h_p- \bar{h}_p\right) \left(\sigma ^2+1 \right) -2 \left(h_p^2+\bar{h}_p^2\right) \right)\right), & \frac{h_w}{c} \ll \eta\\
     4^{h_p+\bar{h}_p} \left(\frac{-(\sigma+i)^2 e^{2\alpha t} }{\epsilon_{12} \epsilon_{34}}\right)^{h_p-\bar{h}_p}  \left( 1+\frac{6 \pi i h_w (\sigma+i)^2}{c \epsilon_{12} \epsilon_{34}}e^{2\alpha t} \right)^{-(2h_v+h_p)}, & \frac{h_w}{c} \sim \eta\\
      16^{h_p} \left(\frac{i \, c}{24\pi h_w}\right)^{2h_v+h_p} \left(-\frac{4 \epsilon_{12} \epsilon_{34}e^{-2 \alpha  t} }{(\sigma+i )^2} \right)^{2h_v+\bar{h}_p}, & \frac{h_w}{c} \gg \eta.
    \end{cases}
\end{align}
The OTOC decays to zero at late times and the time at which it becomes ${\cal O}(1)$ is called the scrambling time,
\begin{align}
    t_* = \frac{(2h_v+h_p)}{2\alpha(2h_v+\bar{h}_p)}\log\left(\frac{c}{h_w}\right).
\end{align}

To understand the above equation better, let's consider the contribution to the OTOC of zero-twist intermediate states, $h_p=\bar{h}_p,$ the OTOC is,
\begin{align}
    {\cal V}_{h_p}(\eta) {\cal V}_{\bar{h}_p}(\bar{\eta}) \approx \begin{cases}
    4^{2h_p}  , & \frac{h_w}{c} \ll \eta\\
     4^{2h_p}   \left( 1-\frac{6 \pi i h_w (2h_v+h_p) (\sigma+i)^2}{c \epsilon_{12} \epsilon_{34}}e^{2\alpha t} \right), & \frac{h_w}{c} \sim \eta\\
      16^{h_p}  \left(-\frac{4 i c\, \epsilon_{12} \epsilon_{34}e^{-2 \alpha  t} }{6\pi h_w (\sigma+i )^2} \right)^{2h_v+h_p}, & \frac{h_w}{c} \gg \eta.
    \end{cases}
\end{align}
Substituting the above equation in \eqref{block_expansion} and \eqref{light_block}, we find that for times greater than dissipation time and less than scrambling time, $\frac{h_w}{c} \sim \eta$, the OTOC indeed decays exponentially with an exponent $2\alpha$. Therefore, the OTOC saturates the bound $\Lambda_L \leq 2\alpha$ also when exchanges of light intermediate states are taken into account for computing the four-point function.

\subsection{Evolution under other unitaries}
In this sub-section we will study the OTOC for growth of product of operators under the action of the unitary, 
\begin{align} \label{unitaryk}
    {U}_k = e^{i \alpha (L_k+L_{-k})t}, \qquad k = \lbrace 2,3, \dots \rbrace.
\end{align}
We will use the fact that $L_{k,-k}, L_0$ operators, which obey Virasoro algebra form a closed algebra, which gives the $SL(2,R)$ algebra on redefining the operators.

Given the Virasoro algebra,
\begin{align}
    \left[ L_n,L_m \right] = (n-m) L_{n+m} + \frac{c}{12} n^2(n^2-1) \delta_{n+m,0}
\end{align}
we analyse a subset of three operators, $L_0$ and $L_{k,-k}$ where $k = \lbrace 2,3, \dots \rbrace$. The following redefinition of these operators,
\begin{align} \label{Ltilde}
    \tilde{L}_0 = \frac{1}{k} \left(L_0 + \frac{c}{24}(k^2-1) \right), \quad \tilde{L}_{\pm} = \frac{1}{k} L_{\pm k}
\end{align}
forms an $SL(2,R)$ algebra. The second term, strikingly, has the same form as the conformal dimension of a $k$-cycle twist operator in orbifold CFTs. This isn’t a coincidence and we will elaborate on this point further in section \ref{sec:modular_hamiltonian}.

The unitary \eqref{unitaryk}, whose $2\times2$ representation is given in equation \eqref{uk22}, acts on the operators as an $SL(2,R)$ transformation and results in a M\"{o}bius transformation of the location of the operators (see derivation in appendix \ref{app:sl2r}, equation \eqref{z_mobius_uk}),
\begin{align}
    z' = \frac{z \cosh(k\alpha t) + i \sinh (k\alpha t) }{\cosh(k\alpha t) - i z \sinh (k\alpha t) }.
\end{align}

The growth of a product of operators under evolution by the unitary of equation \eqref{unitaryk} is the following,
\begin{align}
     U_k(t) V(z_3) U^\dagger_k(t) U_k(t) V(z_4)U^\dagger_k(t) |0 \rangle = \left( \frac{\partial z'}{\partial z} \right)^{h_v}\bigg|_{z=z_3} \left( \frac{\partial z'}{\partial z} \right)^{h_v}\bigg|_{z=z_4} V(z'_3) V(z'_4) |0 \rangle.
\end{align}
Using the above equation and following steps similar to those in equations \eqref{4pt_u1} to \eqref{otoc_u1}, with $\alpha \to \alpha k$, we obtain the following OTOC,
\begin{align} \label{otoc_uk}
    {\cal C} \approx \left( 1-\frac{12\pi i h_w h_v }{c \, \epsilon_{12}\epsilon_{34}}(\sigma+i)^2  e^{2 k \alpha  t} \right).
\end{align}
It was shown in \cite{Caputa:2021sib} that the Krylov complexity for evolution by the unitary ${U}_k$ grows as,
\begin{align} \label{krylov_k}
    K^k_{O} \sim e^{2\alpha k t}.
\end{align}
Therefore, we find that in case of operator evolution by the unitary of equation \eqref{unitaryk} composed of Virasoro generators $L_{\pm k}$, the OTOC decays with a Lyapunov exponent which saturates the bound set by the $K$-complexity growth rate.\footnote{\cite{Parker:2018yvk} proposed that $K$-complexity serves as an upper bound for all $q$-complexities, an OTOC is an example of the latter. While they conjecture this for Hamiltonian evolution, the form of the Hamiltonian is not specified. In addition, evolution under Hamiltonians of the form $H = \alpha L_0 + \gamma (L_k+L_{-k})$ are studied in certain quench protocols such as the SSD quench \cite{Calabrese:2016xau,Fan:2019upv}.}

Here, the exponential term in the OTOC in equation \eqref{otoc_uk} and Krylov complexity in \eqref{krylov_k} has a factor of $2\alpha k$. It is interesting to note that this exponent is the square-root of the Cartan-Killing form of $ U_k$ which is $4\alpha^2k^2$.

\section{Virasoro generators as Modular Hamiltonian} \label{sec:modular_hamiltonian}
In the previous section, we have seen that the OTOC decays as an exponential function of time. It is natural to wonder why we see exponential chaos in a zero temperature CFT. In this section we will show that this behaviour comes about because the operators are evolved with ${ H}_k$, which is the modular Hamiltonian on an arc, for a spatially compact CFT. Since evolution by the modular Hamiltonian gives rise to a thermal system \cite{Casini:2011kv,Cardy:2016fqc}, hence we find that the theory becomes chaotic. 

Here, we review some essential properties of the modular Hamiltonian. For a more complete and recent review see \cite{Dalmonte:2022rlo}. Consider a state $|\phi\rangle$ in a Hilbert space $\cal{H}$ along with a set of observables $\cal{M}$ within a region $A$. The Hilbert space is partitioned into degrees of freedom in the region $A$ and the rest of the system,
\begin{align}
     {\cal H} = {\cal H}_A \otimes {\cal H}_{A^c}.
\end{align}
The reduced density matrices in such a system are described as,
 \begin{align}
     \rho_A  = \Tr_{A^c}|\phi\rangle \langle \phi| \nonumber,
 \end{align}
and similarly for $\rho_{A_c}$. Since the reduced density matrix is Hermitian and positive semi-definite, it can be written as an exponential of a Hermitian operator, $H_{\text{mod}}$,\footnote{Here, $H_{\text{mod}}$ would depend on the state $|\phi\rangle$ and subregion $A$ such that $H_{\text{mod}} \equiv H^{\phi}_A$, however we are suppressing this dependence in this paper.}
\begin{align} \label{density_mod}
    \rho_A = e^{-H_{\text{mod}}}
\end{align}
such that
\begin{align}
    H_{\text{mod}} |\phi \rangle = 0.
\end{align}

The unitary operator, 
\begin{align} \label{Hmod_unitary}
    U(s) = \rho_A^{is} = e^{-is H_{\text{mod}}}
\end{align}
generates a symmetry of the system. If $H_{\text{mod}}$ is a local operator, $U(s)$ generates a local flow in the causal development of the region $A$, also called \textit{modular flow}. Modular flow can be thought of as providing a natural notion of an internal clock for the operator on which $U(s)$ acts and the modular flow parameter, $s$ can be thought of as a notion of time in this internal clock. In this paper, we work with local modular Hamiltonian, $H_{\text{mod}}$, for which there exists a notion of trace defined with respect to the reduced density matrix on the subspace $A$.
Hence for any operator ${\cal O}$ inside the region $A$, 
\begin{align}
    \Tr \left(\rho_A U(s) {\cal O} U(-s) \right) = \Tr \left( \rho_A {\cal O}\right).
\end{align}
Extending the transformations in equation \eqref{Hmod_unitary} to complex parameters, the correlators obey Kubo-Martin-Schwinger (KMS) conditions, which is a relation of periodicity in imaginary time,
\begin{align} \label{kms}
    \Tr \left(\rho_A {\cal O}_1(i) {\cal O}_2 \right) = \Tr \left(\rho_A U(i) {\cal O}_1 U(-i) {\cal O}_2 \right) = \Tr \left( \rho_A {\cal O}_2 {\cal O}_1\right)
\end{align}
where, ${\cal O}(s)$ is the transformation of the operator ${\cal O}$ under the modular flow \eqref{Hmod_unitary},
\begin{align}
    {\cal O}(s) = U(s) {\cal O} U(-s).
\end{align}

\subsection{Modular Hamiltonian on half circle}
In this section, let's have a more careful look at the unitary,
\begin{align} \label{Hmod_1}
    {H}_1 = \alpha (L_{1} + L_{-1}).
\end{align} Substituting the Virasoro generators as an integral over the stress tensor times a power-law in $z$, 
\begin{align}\label{vir_gen_int}
    L_k=\frac{1}{2\pi i} \oint dz z^{k+1} T(z)
\end{align} \label{H1integral}
in equation \eqref{Hmod_1}  
\begin{align}
    {H}_1 = \frac{\alpha}{2\pi i} \oint dz (z^2 + 1) T(z)
\end{align}

Given a $2d$ CFT with compact spatial direction of length $L$ and a finite entanglement interval, $A = (-R,R)$, the modular Hamiltonian is \cite{Cardy:2016fqc},
\begin{align} \label{HmodCardy}
    H_{\text{mod}} = \frac{L}{\pi} \int_A dx \frac{\sin[\frac{\pi(R-x)}{L}] \sin[\frac{\pi(R+x)}{L}]}{\sin[\frac{2\pi R}{L}]}T_{00}(x).
\end{align}
It was shown in \cite{deBoer:2021zlm} that the Hamiltonian ${H}_1 = \alpha (L_{1} + L_{-1})$ is proportional to the modular Hamiltonian on the half circle $x = [-\pi/2,\pi/2]$ for a CFT with compact spatial direction of period $2\pi$. We briefly review the analysis here before generalising it for the unitaries ${H}_k = \alpha (L_{k} + L_{-k})$ in the next subsection. To see this, let's substitute $L=2\pi, R=\pi/2$,
\begin{align}
    z  & = e^{ix}, \quad \text{and}\\
    T_{00}(x) & = -(T(x)+\bar{T}(x))
\end{align}
in equation \eqref{HmodCardy},
\begin{align}
    H_{\text{mod}} = - \int_{-\pi/2}^{\pi/2} dx \cos(x)(T(x)+\bar{T}(x)).
\end{align}
Substituting, $z=e^{i\theta}$, the stress tensor transforms as,
\begin{align}
    T(x) = \left(\frac{\partial z}{\partial x}\right)^2 T(z) + \frac{c}{12} \left\lbrace z,x \right\rbrace
\end{align}
we obtain,
\begin{align} \label{completeHmod}
    H_{\text{mod}} = - \frac{1}{2i} \oint_{|z|=1} dz (z^2+1) T(z)
\end{align}
where in the last step we have ignored the anti-holomorphic part of the stress tensor and subtracted the vacuum energy of the cylinder by ignoring the Schwarzian term. Also note that in equation \eqref{completeHmod}, we are integrating over the full circle instead of $z=(-i,i)$, this is because the two integrals differ only by an overall phase factor. 
Comparing equations \eqref{H1integral} and \eqref{completeHmod}, we find that the former is the modular Hamiltonian up-to an overall factor,
\begin{align} \label{H1tomod}
    {H}_1 = \frac{\alpha}{\pi} H_{mod}.
\end{align}
The modular Hamiltonian for a half-circle given in equation \eqref{completeHmod} annihilates the CFT vacuum.

\subsection{Modular Hamiltonian of an orbifold CFT}
Let's consider the Hamiltonian ${H}_k$ from equation \eqref{Hk} for integers $k>1$. Substituting the Virasoro generators \eqref{vir_gen_int}, it is written as,
\begin{align}
    {H}_k = \frac{\alpha}{2\pi i} \oint dz \left(z^{k+1}+z^{-k+1} \right) T(z).
\end{align}
Performing the following co-ordinate transformation,
\begin{align}\label{transf_z_orb}
    z = e^{i \frac{x}{k}}
\end{align}
we find that 
\begin{align} \label{Hktomod}
    {H}_k = \frac{\alpha k}{\pi} H_{mod,k}.
\end{align}
The co-ordinate transformation of equation \eqref{transf_z_orb} creates a $k$-sheeted Riemann surface which is a cyclic orbifold, with cycle length $k$.

Comparing equation \eqref{Hktomod} with equation \eqref{HmodCardy},
\begin{align}
    H_{mod,k} = \int_{-k\pi/2}^{k\pi/2} dx \cos x \, T_{00}(x) 
\end{align}
is the modular Hamiltonian of the interval $A = [-R,R]$ where $R=\frac{k\pi}{2}$ in a CFT with compact spatial direction of length $L=2\pi k$. Half of this CFT is traced out to obtain the above modular Hamiltonian. Here, again we have converted the integral over the full circle to an integral over half circle, since the two differ by an overall phase factor, which we ignore. This is the modular Hamiltonian of a half-interval, of an orbifold CFT whose vacuum energy is shifted by $\frac{c(k^2-1)}{24k}$ which is the second term in the transformation from $L_0 \to {\tilde L}_0$ in equation \eqref{Ltilde}.\footnote{The quantity $\frac{c(k^2-1)}{24k}$ is the ground state energy of the $k$-cycle twisted sector and thus the conformal dimension of the twist operator.} $H_{mod,k}$ annihilates the ground state of the orbifold CFT.

\section{Operator growth bound from chaos bound} \label{sec:bound}
In the previous section we have seen that the Hamiltonians of the form ${H}_k$ given by equation \eqref{Hk}, where $k$ is a positive integer, are related to the modular Hamiltonian (equation \eqref{Hktomod}) of a CFT with half its spatial region traced out. 
It is known that the modular Hamiltonian, obtained by tracing out half space of the CFT and which annihilates the CFT vacuum, gives rise to a local modular evolution \cite{Balakrishnan:2017bjg}. Therefore, evolution by these Hamiltonians ${H}_1, {H}_k$ in a zero temperature $2d$ CFT, gives rise to thermal dynamics. In \cite{Maldacena:2015waa}, a bound was proposed on the growth of the OTOC in a thermal quantum system. In this section, we generalise the arguments of \cite{Maldacena:2015waa} to derive a bound on the growth of the OTOC under evolution by the unitary \eqref{unitary1}.

Under evolution by the modular Hamiltonian of half space, the CFT admits a thermal description with density matrix given in equation \eqref{density_mod}. The correlation function of operators evolved by a local modular Hamiltonian satisfy the KMS conditions given in equation \eqref{kms}. Using this property, it can be shown that the two point function, ${\cal B}(\tau)=\Tr \left[\rho_A^{-1/2} W(\tau) \rho_A^{-1/2} W(\tau) \right]$, of some Hermitian operator $W$ within the entanglement interval, is analytic in the half-strip, $-\frac{\pi}{2\alpha} \leq \tau \leq \frac{\pi}{2\alpha}$.

It was shown in \cite{Maldacena:2015waa} that the OTOC in a thermal system may be written as,
\begin{align} \label{OTOC_formal}
    F(t) = \Tr \left[y W y V(t) y W y V(t) \right]
\end{align}
where
\begin{align}
    y^4 \equiv  \rho_A = e^{-H_{mod}}.
\end{align}
The operators are out-of-time ordered on the imaginary time circle of period $\frac{\pi}{\alpha}$, each of them separated by a distance $\frac{\pi}{4\alpha}$. The operators $W,V$ are located within the entanglement interval $A$. Since, the real part of complex co-ordinate $z$ always decreases under M\"{o}bius transformation in equations \eqref{z_mobius_u1} and \eqref{z_mobius_uk}. Therefore, if we start with an operator inside the half-space, it remains inside the half-space under evolution by the Hamiltonian of equation  \eqref{unitaryk}.

We define the normalised OTOC,  
\begin{align} \label{OTOC_formal_norm}
    f(t) & = \frac{F(t)}{F_d} = \frac{\Tr \left[y W y V(t) y W y V(t) \right]}{\Tr \left[y^4 WW \right] \Tr \left[y^4 V(t) V(t) \right]}, 
\end{align}
where, $F_d = \Tr \left[y^4 WW \right] \Tr \left[y^4 V(t) V(t) \right]$ is the disconnected term composed of two point functions of $VV$ and $WW$ operators. If the normalised OTOC satisfies the following,
\begin{itemize}
    \item $f(t+i \tau)$ is an analytic function in the region $0<t$ and $-\frac{\pi}{4\alpha} \leq \tau \leq \frac{\pi}{4\alpha}$,
    
    \item $|f(t+i\tau)| \leq 1$ in the entire half strip,
\end{itemize}
then $f(t)$ satisfies the following,
\begin{align}
    \frac{1}{1-f} \left| \frac{df}{dt} \right| \leq 2\alpha + O(e^{-4\alpha t}).
\end{align}
Substituting the following form of the function, $f(t) = 1- \epsilon e^{\Lambda_L t}$, gives the bound on Lyapunov exponent,
\begin{align}
    \Lambda_L \leq 2\alpha.
\end{align}
The above two properties may be shown by following arguments similar to those in the paper \cite{Maldacena:2015waa}, where the temperature of this thermal system is $\beta \to \frac{\pi}{\alpha}$. For the sake of completeness, we include these arguments in the appendix \ref{app:bound}

Similar considerations as the derivation of the chaos bound were used in \cite{Balakrishnan:2017bjg} to obtain the quantum null energy condition. In \cite{DeBoer:2019kdj} evolution of a CFT subspace by the modular Hamiltonian was considered, furthermore the modular Hamiltonian was perturbed, $\delta H_{mod}$. A bound on modular chaos was proposed by studying the expectation value of the modular evolution of this modular Hamiltonian perturbation. It was shown that the expectation value of modular evolution of $\delta H_{mod}$ grows exponentially with an exponent $\lambda$ which is bounded, $\lambda \leq 2\pi$.

\section{Conclusion}
We computed the OTOC in a $2d$ CFT when the operators grow under evolution by the Hamiltonian in equation \eqref{Hk}. We found that the Lyapunov exponent saturates the bound $\Lambda_L \leq 2 \alpha$ proposed in \cite{Parker:2018yvk}. This indicates that a holographic dual exists and is a black-hole with inverse temperature $\beta = \frac{\pi}{\alpha}$. Driving by Hamiltonians ${ H}_{k}$ is considered in Floquet CFTs \cite{Fan:2019upv, Das:2021gts}, which has connections to Floquet dynamics in quantum mechanical systems and lattice models. Therefore, the existence of this holographic dual could provide a way towards bridging the gap between traditional methods of measuring quantum information scrambling in out-of-equilibrium quantum systems and the methods proposed in the context of black-holes.

OTOC in CFTs which are periodically driven by a combination of the unitary \eqref{unitary1} and the usual CFT Hamiltonian $\propto L_0$ has been studied \cite{Das:2022jrr}. However, evolving the CFT by the unitary ${H}_k$ alone provided a setup to observe thermal behaviour arise in a zero temperature CFT. The latter follows from the observation that ${H}_k$ is a modular Hamiltonian with half-space traced out. Another way to obtain thermal OTOC in a zero temperature CFT is to consider the six point function of two heavy and four light operators, where the conformal dimension of the heavy operator is related to the effective temperature \cite{Anous:2019yku}. 

The entanglement entropy following evolution of states in a CFT under evolution by a combination of Virasoro generators $L_{\pm k}$ was studied in \cite{Caputa:2022zsr}. Furthermore, they propose the holographic dual of such a CFT to be Ba$\tilde{\text{n}}$ados geometry. It will be interesting to compute the OTOC in this dual geometry. 

For CFTs undergoing quench, understanding whether the OTOC satisfies the bound on Lyapunov exponent proposed in \cite{Maldacena:2015waa} has been a topic of several recent studies \cite{David:2017eno,David:2019bmi,Khetrapal:2018jyn}. However it is not well understood how the bound of \cite{Maldacena:2015waa} would change for a CFT undergoing quench. \cite{Cardy:2016fqc} showed that a CFT undergoing quench can be represented as an evolution by a modular Hamiltonian. Using the methods of sections \ref{sec:modular_hamiltonian} and \ref{sec:bound}, it would be interesting to understand how the bound on Lyapunov exponent could be derived for these time dependent geometries.

\subsection*{Comments on subleading bounds and extremal OTOC}
In \cite{Kundu:2021qcx,Kundu:2021mex}, it was argued that the OTOC satisfies infinitely many constraints obtained by imposing positivity and boundedness, monotonicity and log-convexity on its moments. This infinite set of constraints on the OTOC includes the chaos bound $\lambda_L \leq \frac{2\pi}{\beta}$ as the leading constraint. Consider an OTOC, $f(t)$, of the form, $1 -f(t) \sim  a_1 e^{\lambda_L t} + 
\epsilon a_2 e^{\lambda_2 t} + \dots$, where $a_1, a_2$ are ${\cal O}(\epsilon^0)$ coefficients and $\epsilon >0 $ is a small dimensionless parameter such that the second term is sub-leading. Then the infinite set of constraints impose a bound on the growth of the subleading term, $\lambda_2 \leq \frac{6\pi}{\beta}$, when the leading term saturates the above bound. Similarly, bounds are obtained for $\lambda_3$ when both $\lambda_1=\lambda_L$ and $\lambda_2$ saturate the above bounds. It is expected that it would be possible to extend the results of section \ref{sec:bound} such that the thermal OTOC with inverse temperature $\beta = \frac{\pi}{\alpha}$ would satisfy the subleading bounds,
\begin{align}
    \Lambda_1 \leq 2\alpha, \quad \Lambda_2 \leq 6\alpha, \quad \dots \quad \Lambda_n \leq 2\alpha(n-1).
\end{align}
Here, $\Lambda_i$ are the exponents of the terms in the OTOC when the theory is thermal with inverse temperature $\pi/\alpha$. This provides further sub-leading bounds which are set on the OTOC by operator growth. Furthermore, the extremal OTOC which satisfies all the above bounds is defined to be,
\begin{align}
   f_{ext}(t) = 1 - a_1 {\cal F}_{ext}(t,t_{eff}),
\end{align}
where,
\begin{align}
    {\cal F}_{ext}(t,t_{eff}) = \frac{e^{2\alpha t}}{1+e^{4\alpha(t-t_{eff})}}.
\end{align}
Here, $t_{eff}$ is the time at which the extremal OTOC has a minima. In \cite{Kundu:2021mex}, a K\"{a}ll\'{e}n-Lehmann type representation of the extremal OTOC was proposed,
\begin{align}
    1-f_{ext}(t) = \int_{t_0}^\infty ds \, {\cal F}_{ext} (t,s) \rho(s), \quad 0\leq \rho(s)\leq \frac{8\alpha}{\pi} e^{-2\alpha s} .
\end{align}
We anticipate that the $K$-complexity could be derived from the extremal OTOC with appropriate choice of the density function, $\rho$. In this way the extremal OTOC would encompass measures of chaos such as the OTOC and $K$-complexity.

 \paragraph{Acknowledgements} We would like to thank Shouvik Datta for collaboration during the early stages of this work, for helpful discussions throughout and for comments on the draft. We thank Justin David, Pawel Caputa, Diptarka Das, R. Loganayagam and Subham Dutta Chowdhury for discussion and comments. We thank IIT Mandi for their hospitality during the workshop ``Out of equilibrium Physics" where part of the results in this paper were presented. SK's research is supported by Department of Science and Technology's INSPIRE grant DST/INSPIRE/04/2020/001063.

\appendix

\section{Co-ordinate transformation under operator evolution} \label{app:sl2r}
In this appendix, we derive the transformation of an operator under the action of the unitary,
\begin{align}
    {U}_1(t) = e^{i\alpha(L_{1}+L_{-1})t}.
\end{align}
The action of the above unitary on a single operator in Euclidean space is,
\begin{align}
    V(t) = {U}_1(t)V(z){U}_1^\dagger(t).
\end{align}
This acts as an $SL(2,R)$ transformation,
\begin{align} \label{u122}
    {U}_1(t) = \left(
\begin{array}{cc}
 \cosh (\alpha  t) & i \sinh (\alpha  t) \\
 -i \sinh (\alpha  t) & \cosh (\alpha  t) \\
\end{array}
\right)
\end{align}
where we have used the $2\times2$ representation of the $SL(2,R)$ matrices,
\begin{align}
    L_0 = \left(
\begin{array}{cc}
 \frac{1}{2} & 0 \\
 0 & -\frac{1}{2} \\
\end{array}
\right), \quad L_{1} = \left(
\begin{array}{cc}
 0 & 0 \\
 -1 & 0 \\
\end{array}
\right), \quad L_{-1} = \left(
\begin{array}{cc}
 0 & 1 \\
 0 & 0 \\
\end{array}
\right).
\end{align}
Thus, under the $SL(2,R)$ transformation, the location $z$ of the operator $V(z)$ undergoes M\"{o}bius transformation,
\begin{align}
    z \to z' = \frac{a z+b}{c z+d}
\end{align}
where,
\begin{align}
    \left(
\begin{array}{cc}
 a & b \\
 c & d \\
\end{array}
\right) = \left(
\begin{array}{cc}
 \cosh (\alpha  t) & i \sinh (\alpha  t) \\
 -i \sinh (\alpha  t) & \cosh (\alpha  t) \\
\end{array}
\right).
\end{align}
Therefore, we obtain 
\begin{align} \label{z_mobius_u1}
    z' = \frac{z \cosh (\alpha  t)+i \sinh (\alpha  t)}{\cosh (\alpha  t)-i z \sinh (\alpha  t)}.
\end{align}

We can do a similar analysis for evolution of an operator by the unitary ${U}_k$ defined in equation \eqref{unitaryk}. We have seen that on redefining $L_{\pm k}$ in terms of ${\tilde L}_{\pm}$ in equation \eqref{Ltilde}, ${\tilde L}_{0,\pm k}$ form an $SL(2,R)$ algebra. Using the $2\times2$ representation for the $SL(2,R)$ matrices ${\tilde L}_{0,\pm k}$, we find ${U}_k$ is,
\begin{align} \label{uk22}
    {U}_k(t) = \left(
\begin{array}{cc}
 \cosh (\alpha k t) & i \sinh (\alpha k t) \\
 -i \sinh (\alpha k t) & \cosh (\alpha k t) \\
\end{array}
\right).
\end{align}
Therefore, under the action of this unitary, an operator transforms as, $V(z') = {U}_k(t) V(z) {U}_k^\dagger(t)$, where 
\begin{align} \label{z_mobius_uk}
    z \to z' = \frac{z \cosh (\alpha k t)+i \sinh (\alpha k t)}{\cosh (\alpha k t)-i z \sinh (\alpha k t)}.
\end{align}

\section{OTOC in the LLLL limit}
In this appendix, we derive the OTOC in the light-light-light-light (LLLL) limit, i.e. when $h_v,h_w,c \to \infty$ and $h_v/c \to 0, h_w/c \to 0$ with $h_v h_w/c$ held fixed \cite{Fitzpatrick:2014vua}. In this limit, the Virasoro identity block exponentiates,
\begin{align}
    {\cal V}_0 = \exp \left[ 2\frac{h_v h_w}{c} \eta^2 \, _2F_1 (2,2,4;\eta)\right]
\end{align}
where, 
\begin{align}
     _2F_1 (2,2,4;\eta) = \frac{6 (-2 \eta +\eta  \log (1-\eta )-2 \log (1-\eta ))}{\eta ^3},
\end{align}
which has a branch cut along $\eta = [0,\infty)$. To obtain the OTOC, we continue the cross-ratio across the branch cut, as in equation \eqref{eta_continue}, followed by taking $\eta \to 0$ limit.

In order to obtain the OTOC, we only need the contribution to leading order in the expansion in small $\frac{h_w h_v}{c}$. The OTOC in this limit is,
\begin{align}
    {\cal C} & = 1 + \frac{2h_v h_w}{C} \eta^2 \, _2F_1(2,2,4;\eta) + \dots\\
    & = 1+\frac{48 h_w h_v \pi i}{c \eta} + \dots,
\end{align}
where the second line is obtained by performing the analytic continuation of $\eta$ followed by taking the small $\eta$ limit. Substituting the cross-ratio from equation \eqref{cross-ratio}, we obtain the OTOC,
\begin{align}
    {\cal C} = 1-\frac{12 h_w h_v \pi i (\sigma+i)^2}{c \epsilon_{12} \epsilon_{34}} e^{2\alpha t} + \dots.
\end{align}

\section{Modular Hamiltonian to CFT Hamiltonian}
In this appendix, we show a co-ordinate transformation which takes the modular Hamiltonian composed of Virasoro generators $L_{\pm 1}$ to the usual CFT Hamiltonian $\propto L_0$.\footnote{We thank Diptarka Das for suggesting this computation.}
Given the unitary,
\begin{align}
    {H}_1 = \frac{\alpha}{2\pi i}\int_{-i}^i (1+z^2) T(z) dz,
\end{align}
let us substitute a transformation, $z \to f(\tilde{z})$ such that,
\begin{align}
    {H}_1 = \frac{\alpha}{2\pi i}\int \tilde{z} T(\tilde{z}) d\tilde{z}.
\end{align}
Here the range of integral is to be determined from the transformation $f(\tilde{z})$ and $T(\tilde{z})$ transforms as,
\begin{align}
    T(z) = \left( \frac{\partial \tilde{z}}{\partial z} \right)^2 T(\tilde{z}) + \frac{c}{12} \left\lbrace \tilde{z},z \right\rbrace .
\end{align}
Ignoring the Schwarzian term, we obtain the unitary to be,
\begin{align}
    {H}_1 = \frac{\alpha}{2\pi i}\int \frac{1+f(\tilde{z})^2}{f'(\tilde{z})} T(\tilde{z}) d\tilde{z},
\end{align}
and $f(\tilde{z})$ is obtained by imposing,
\begin{align}
    \frac{1+f(\tilde{z})^2}{f'(\tilde{z})} = \tilde{z}.
\end{align}
Solving the above equation we obtain,
\begin{align}
    \tilde{z} = \exp \left( \tan^{-1}\left(z\right) -\frac{\pi}{4}\right)
\end{align}
Substituting $z=f(\tilde{z}) = e^{i\theta}$, for $\theta = (-\frac{\pi}{2},\frac{\pi}{2})$, $|\tilde{z}| = 1$, therefore the unitary becomes,
\begin{align}
    {H}_1 & = \frac{\alpha}{2\pi i}\oint \tilde{z} T(\tilde{z}) d\tilde{z}\\
    & = \alpha L_0,
\end{align}
in the new co-ordinate $\tilde{z}$. Thus the evolution by the unitary can be understood to be the usual CFT time evolution in this new system with co-ordinate $\tilde{z}$ and temperature $\frac{\pi}{\alpha}$.

\section{Analyticity and boundedness of OTOC} \label{app:bound}
In this appendix, we show that the OTOC in equation \eqref{OTOC_formal_norm} satisfies the analyticity and boundedness properties given in section \ref{sec:bound}. For the first property, note that equation \eqref{OTOC_formal_norm} for $t+i\tau$ can be written as,
\begin{align}
    f(t +i \tau) = \frac{1}{F_d} \Tr \left( e^{-\left(\frac{\pi}{4\alpha}-\tau\right)} W e^{-\left(\frac{\pi}{4\alpha}+\tau\right)} V(t) e^{-\left(\frac{\pi}{4\alpha}-\tau\right)} W e^{-\left(\frac{\pi}{4\alpha}+\tau\right)} V(t) \right).
\end{align}
The above function diverges for $\tau > \frac{\pi}{4\alpha}$ and $\tau < -\frac{\pi}{4\alpha}$, however it is convergent for  $ - \frac{\pi}{4\alpha} \leq \tau  \leq \frac{\pi}{4\alpha}$. In addition, for a functional form $f(t) = 1- \epsilon e^{\Lambda_L t}$, it is infinitely differentiable. Therefore, we may conclude that $F(t)$ is an analytic function in the region $ - \frac{\pi}{4\alpha} \leq \tau  \leq \frac{\pi}{4\alpha}$.

The second property of boundedness can be shown upto an error of $\epsilon$,
\begin{align}
    & f(t)  = \frac{F(t)}{F_d + \epsilon},\\
   & |f(t+i\tau)|  \leq 1.
\end{align}
This is done by first showing that $|F(t+i\tau)| \leq F_d+\epsilon$ at all the three boundaries $\tau = \pm \frac{\pi}{4\alpha}$ and at $t = t_0 > 0$, where $t_0$ is small value approaching $0$. At the boundary $\tau = \pm \frac{\pi}{4\alpha}$, the OTOC is
\begin{align} \label{bdy_pm}
    F \left(t\pm i \frac{\pi}{4\alpha} \right) & = \Tr \left[y^2 W V(t) y^2 W V(t) \right], \nonumber \\
    |F \left(t\pm i \frac{\pi}{4\alpha} \right)| &\leq \Tr \left[y^2 V(t) W y^2 W V(t)  \right], \nonumber \\
    & \leq F_d + \epsilon,
\end{align}
where in the second line the Cauchy-Schwarz inequality is used. For a chaotic system, at $t \gg t_d$ where $t_d$ is the dissipation time, the correlator in the second line factorises to $F_d$ and $\epsilon$ takes into account the possible error in factorisation. Let's now consider the OTOC, $F(t_0+i\tau)$, on the boundary $t = t_0 \gtrsim 0$, where the time $t_0 \ll t^*$, where $t^*$ is the scrambling time. In this case, this correlator is still time-ordered, therefore
\begin{align} \label{bdy_t0}
    |F \left(t_0+i\tau\right)| & \leq F_d + \epsilon,
\end{align}
where $\epsilon$ takes into account the possible error due to failure of the time-ordered correlator to factorise, or error due to onset of scrambling. Since the function $F(t+i\tau)$ is complex differentiable and its absolute value is bounded at the boundaries of the half-strip by equations \eqref{bdy_pm} and \eqref{bdy_t0}, therefore by the Phragm\'{e}n–Lindelöf principle, its absolute value is bounded in the entire half-strip. Thus proving $|f(t+i\tau)| \leq 1$ is an analytic function in the region $0<t$ and $-\frac{\pi}{4\alpha} \leq \tau \leq \frac{\pi}{4\alpha}$.

\bibliographystyle{JHEP}
\bibliography{references.bib}

\end{document}